\documentclass[8.5pt,twoside,twocolumn]{article}

\usepackage[super,sort&compress,comma]{natbib} 
\usepackage{mhchem}
\usepackage{times,mathptmx}
\usepackage{sectsty}
\usepackage{balance} 

\usepackage{lastpage}
\usepackage[format=plain,justification=raggedright,singlelinecheck=false,font=small,labelfont=bf,labelsep=space]{caption} 
\usepackage{fancyhdr}
\usepackage{epsfig}
\usepackage{epsf,graphicx,amssymb,amsmath,amsthm}
\usepackage{color}
\usepackage{bm}

\definecolor{blue}{rgb}{0,0,1}

\begin{document}

\thispagestyle{plain}
\fancypagestyle{plain}{
\renewcommand{\headrulewidth}{1pt}}
\renewcommand{\thefootnote}{\fnsymbol{footnote}}
\renewcommand\footnoterule{\vspace*{1pt}%
\hrule width 3.4in height 0.4pt \vspace*{5pt}} 
\setcounter{secnumdepth}{5}

\makeatletter 
\def\subsubsection{\@startsection{subsubsection}{3}{10pt}{-1.25ex plus -1ex minus -.1ex}{0ex plus 0ex}{\normalsize\bf}} 
\def\paragraph{\@startsection{paragraph}{4}{10pt}{-1.25ex plus -1ex minus -.1ex}{0ex plus 0ex}{\normalsize\textit}} 
\renewcommand\@biblabel[1]{#1}            
\renewcommand\@makefntext[1]%
{\noindent\makebox[0pt][r]{\@thefnmark\,}#1}
\makeatother 
\renewcommand{\figurename}{\small{Fig.}~}
\sectionfont{\large}
\subsectionfont{\normalsize} 

\fancyfoot{}
\fancyhead{}
\renewcommand{\headrulewidth}{1pt} 
\renewcommand{\footrulewidth}{1pt}
\setlength{\arrayrulewidth}{1pt}
\setlength{\columnsep}{6.5mm}
\setlength\bibsep{1pt}

\twocolumn[
  \begin{@twocolumnfalse}
\noindent\LARGE{\textbf{Phonon gap and localization lengths in floppy materials}}
\vspace{0.6cm}

\noindent\large{\textbf{Gustavo D\"uring, Edan Lerner and Matthieu Wyart$^{\ast}$ }}\vspace{0.5cm}


\noindent\textit{\small{\textbf{Received 16th April 2012}}}

\vspace{0.6cm}

\noindent \normalsize{Gels of semi-flexible polymers, network glasses made  of low valence elements,  softly compressed ellipsoid particles and dense suspensions under flow are examples of floppy materials.
These systems present collective motions  with almost  no restoring force. To study theoretically and numerically the frequency-dependence of the response of these materials,
and  the length scales that characterize their elasticity, we use a model of isotropic floppy elastic networks. 
We show that such networks present a phonon gap for frequencies smaller than a  frequency  $\omega^*$ governed by coordination, and that the elastic response is characterized, and in some cases localized, on a length scale $l_c\sim 1/\sqrt{\omega^*}$ that diverges as the  phonon gap vanishes (with a logarithmic correction in the two dimensional case). $l_c$ also characterizes velocity correlations under shear, whereas another length scale $l^*\sim 1/\omega^*$ controls the effect of pinning boundaries on elasticity.  We discuss the implications of our findings for  suspensions flows, and the correspondence between floppy materials and amorphous solids near unjamming, where $l_c$ and $l^*$ have also been identified but where their roles are not fully understood.}
\vspace{0.5cm}
 \end{@twocolumnfalse}
  ]




\section{Introduction}

\footnotetext{\textit{$^{a}$~New York University, Center of Soft Matter Research, 4 Washington Place, New York, NY,10003,USA. E-mail: mw135@nyu.edu}}

In 1864 Maxwell \cite{maxwell} showed that in order to be mechanically stable, the average coordination $z$ of  structures made of points connected by rigid bars must be larger than a threshold value $z_c$. 
Most common solids  satisfy a microscopic version of this constraint.\cite{alexander,Wyart053} If the constraint is violated, collective modes with no restoring forces (floppy modes) exist. These materials will be referred as   strictly floppy systems. However, in some cases  these modes are stabilized 
by weak interactions:  the  bending energy of semi-flexible polymers confers a finite elasticity to gels,\cite{frey}  as do van der Waals interactions in weakly-coordinated covalent glasses.\cite{
PhillipsThorpe} Floppy modes can also be stabilized by the pre-stress applied on the system, as is the case for gently compressed packings of ellipsoid particles.\cite{Donevsc,mailman,zorana}  Fluids, on the other hand, can display modes that are strictly floppy.  For example, in granular systems or suspensions of hard particles, large and sometimes percolating clusters of connected particles can be formed,\cite{Cheng,
bonamy} and motion within these clusters can only occur along floppy modes  where no particles overlap. As the density increases toward  jamming, the viscosity \cite{jaeger,durian,
boyer} and  the length scale \cite{pouliquen2004,olsson,lespiat} characterizing the correlation of the dynamics diverge, up to the point where floppy modes disappear and the dynamics stops.

In simplified  numerical models of ellipsoid particles,\cite{mailman,zorana}, covalent networks,\cite{silica1}, gels of semi-flexible polymers \cite{Huisman} and  suspension flows \cite{edan} it has been observed that the   density of vibrational modes $D(\omega)$ displays a  gap at low frequency (in addition to the floppy or nearly-floppy modes present at zero-frequency). In suspension flows the amplitude of the gap was shown to affect the divergence of the viscosity near jamming.\cite{edan}  An early work by Garboczi and Thorpe \cite{GarbocziThor1985} supported the idea that a gap of modes exist in floppy materials.   However, the dependence of the gap on the microscopic structure,  and its consequences on the material properties and the different length scales characterizing the elastic response are not understood. 

 In this manuscript, we study the elastic properties of  isotropic harmonic spring networks that are strongly disordered, but where spatial fluctuations of coordination are weak. The model networks we use are presented and analyzed numerically  in section \ref{MD}. In section \ref{EMT}, we show, using both mean-field methods \cite{xiaoming,thorpe2,GarbocziThor1985,wyart2010,mao} and  numerical  simulations, that strictly floppy elastic networks present a vibrational gap between floppy modes (zero frequency modes) and a frequency $\omega^*\sim z_c-z\equiv \delta z$.   The absence of low frequency phonons suppress the  elastic propagation of the  response at frequencies smaller than $\omega^*$. As a consequence, we show  that  the mean  response to a  local perturbation (local strain) is localized,   displaying  an exponential  decay with a characteristic length  $l_c\sim 1/\sqrt \delta z$. we also predict a logarithmic correction to the scaling for  two dimensional systems. In section \ref{BMF} we go beyond mean field and estimate the fluctuations around the mean response. We show that fluctuations dominate the amplitude over the mean value, however, the fluctuation around the mean response also  decays with the same length scale $\sim l_c$. Even though  $l_c$ can be considered to be the localization length of floppy modes, we  show  that these modes can not exist in a region of typical radius smaller than $l^*\sim 1/\delta z$. Surprisingly,  floppy modes can be  be strictly localized (have a compact support) only on the much larger scale $l^*$. In section \ref{WIF}  we study the  floppy networks stabilized by weak interactions, which  confer a finite but small restoring force to the material. We model the weak interaction by adding springs whose stiffnesses is very small.  We find that localization is lost even when  weak interactions have a vanishingly small amplitude, showing  that the response of strictly floppy systems is a singular limit.  However,  $l_c$ still characterizes the response near the applied strain. In  section \ref{RDS} we compare our results with recent observations in the affine solvent model, a simplified model for suspension flows of hard spheres. For this strictly floppy system the spectrum associated to the evolution operator  shares both  commonalities and striking differences with the density of states of the isotropic floppy networks  considered here, leading to a prediction on some dynamical length scale in flow. Finally, our results  raise questions associated with the respective role of $l_c$ and $l^*$ both in floppy materials and in jammed packings, which are discussed in the last section.

\section{Model description and simulation}  
\label{MD}
We consider a network of $N$ point particles of mass $m$, connected by $N_c$ harmonic un-stretched springs in spatial dimension $d$. The elastic energy following a deformation vector field $\{\delta {\bf R}^i\}_{i=1\ldots N}$ is:
\begin{equation}
\label{0}
\delta E=\sum_{ \langle ij\rangle}\frac{k_{ij}}{2}[(\delta {\bf R}_i-\delta {\bf R}_j)\cdot {\bf n}_{ij}]^2+o(\delta {\bf R}^2),
\end{equation}
 where the index $ \langle ij\rangle$ runs over the $N_c$ springs,  whose stiffnesses are $k_{ij}$ and whose directions are along the unit vectors ${\bf n}_{ij}$. Floppy modes correspond to  displacements for which $\delta E=0$. Since the energy is the sum of $N_c$ positive definite terms, it vanishes only if all these terms are zero. Such modes always exist if the number of degrees of freedom $Nd$ is larger than the number of constraints $N_c$, or equivalently if $z\equiv 2N_c/N<z_c=2d$.

 The force field $\vert {\bf F} \rangle$ generated by displacements can be obtained by taking the derivative of Eq.(\ref{0}).  One obtains $\vert {\bf F} \rangle=  \mathcal{M} \vert \delta{\bf R} \rangle$, where:
 \begin{equation}
 \label{00}
 \mathcal{M} =\sum_{\langle ij \rangle}k_{ij}{\bf n}_{ij} \otimes {\bf n}_{ij} (\vert i\rangle-\vert j\rangle)(\langle i\vert-\langle j\vert)
 \end{equation}
 is the stiffness matrix. We have used the bra-ket notation  for which $\langle i\vert \delta {\bf R}\rangle=\delta{\bf R}_i$. The normal modes are given by the eigenvectors of $\mathcal{M}$, and their frequencies are given by the square root of their associated eigenvalues. Herein, we will investigate the spectrum of $\mathcal{M}$ and the spatial properties of its eigenvectors when $z<z_c$.

As a model system we consider isotropic disordered elastic networks, generated by following the method of Ref.\cite{Wyartmaha}: we prepare amorphous packings of compressed soft elastic particles with a coordination significantly larger than $z_c$. The centers of the particles are the nodes of our networks, and the contacts between particles are replaced by un-stretched springs of stiffness $k$. Springs are then removed  until the desired coordination $z$ is reached. Removal takes place randomly from the set of  most connected pairs of nodes, leadings to isotropic networks with low heterogeneity in density and coordination.  Such networks are appropriate models of amorphous solids for which large spatial heterogeneities are not energetically  favorable.\cite{Wyartmaha,wyart2010} In  rigidity percolation,\cite{thorpe2, GarbocziThor1985} spring removal is completely random, resulting in networks with large  fluctuations that affect the elastic properties.

\begin{figure}[htb]
 \begin{center}
 \centerline{ \includegraphics[width=0.5\textwidth]{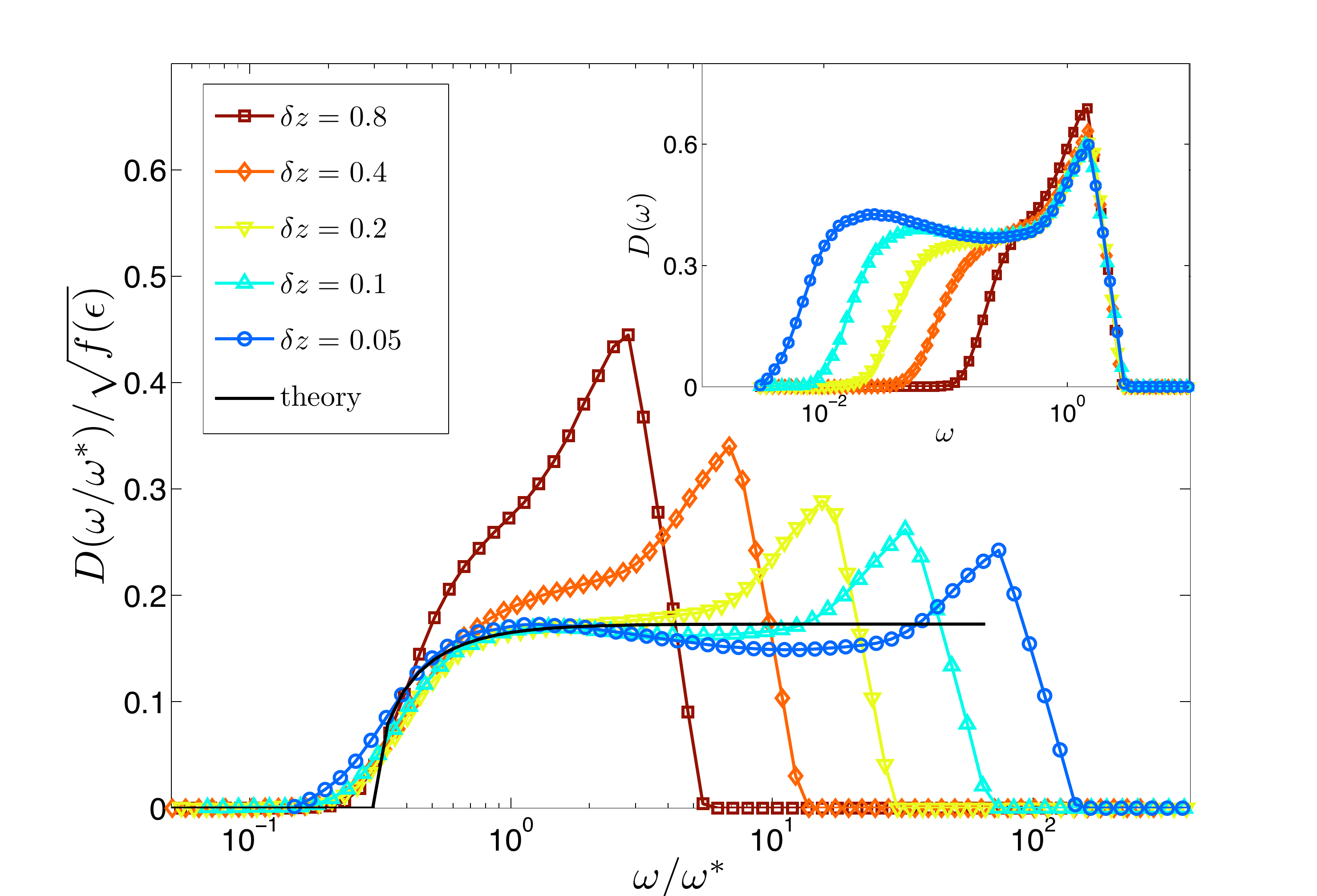} }
 \caption{  \label{DOSFI} {Rescaled density of states {\it vs.} rescaled frequency for $z\in[3.2,3.95] $ using $N=10000$ nodes in two dimensions.  The continuous line corresponds to the theoretical prediction of Eq.(\ref{density}). Inset: Non-rescaled density of states $D(\omega)$ {\it vs} $\omega$. Floppy modes lead to a delta function at $\omega=0$ and are not presented.}}
 \end{center}
 \end{figure}

For these networks, we diagonalize ${\cal M}$ numerically to compute the density of states $D(\omega)$ for various $z<z_c$. As appears in the inset of Fig.\ref{DOSFI}, we find that  a gap in the vibrational spectrum appears below some frequency $\omega^*$ that decreases as $z\rightarrow z_c$. If spatial fluctuations of coordination were large,  one would expect this gap to be filled-up in the thermodynamic limit due to Griffiths-like singularities, but we do not observed this effect in the network considered here.   The role of this gap in elasticity can be analyzed by considering the response to a local strain. We change the rest length of one spring and let the system relax to zero energy under  over-damped dynamics. Such a  response can also be obtained in the absence of damping, by imposing a local oscillatory strain at a vanishing small frequency.  As shown in Fig.\ref{figNonAffine}, the elastic information does not propagate: the response is localized on some length scale that appears to diverge as the gap vanishes, i.e as $z\rightarrow z_c$. The response appears to be very heterogeneous. In what follows we will investigate its mean and its fluctuations.

 \section{Effective medium theory (EMT)}
 \label{EMT}
 \subsection{General formalism} 
 We use EMT, also known as coherent potential approximation,\cite{Kirkpatrick,thorpe2,xiaoming} to investigate the behavior of floppy  networks as the one observed in the previous section.  EMT is a mean field  approximation and neglects spatial fluctuations in the coordination. Therefore, it does not properly describe rigidity percolation.\cite{thorpe2} However, it was shown to describe materials for which  spatial fluctuations of coordination are small.\cite{wyart2010} 
  \begin{figure}[htb]
 \begin{center}
 \centerline{ \includegraphics[width=0.5\textwidth,trim=20mm 0mm 30mm 0mm, clip]{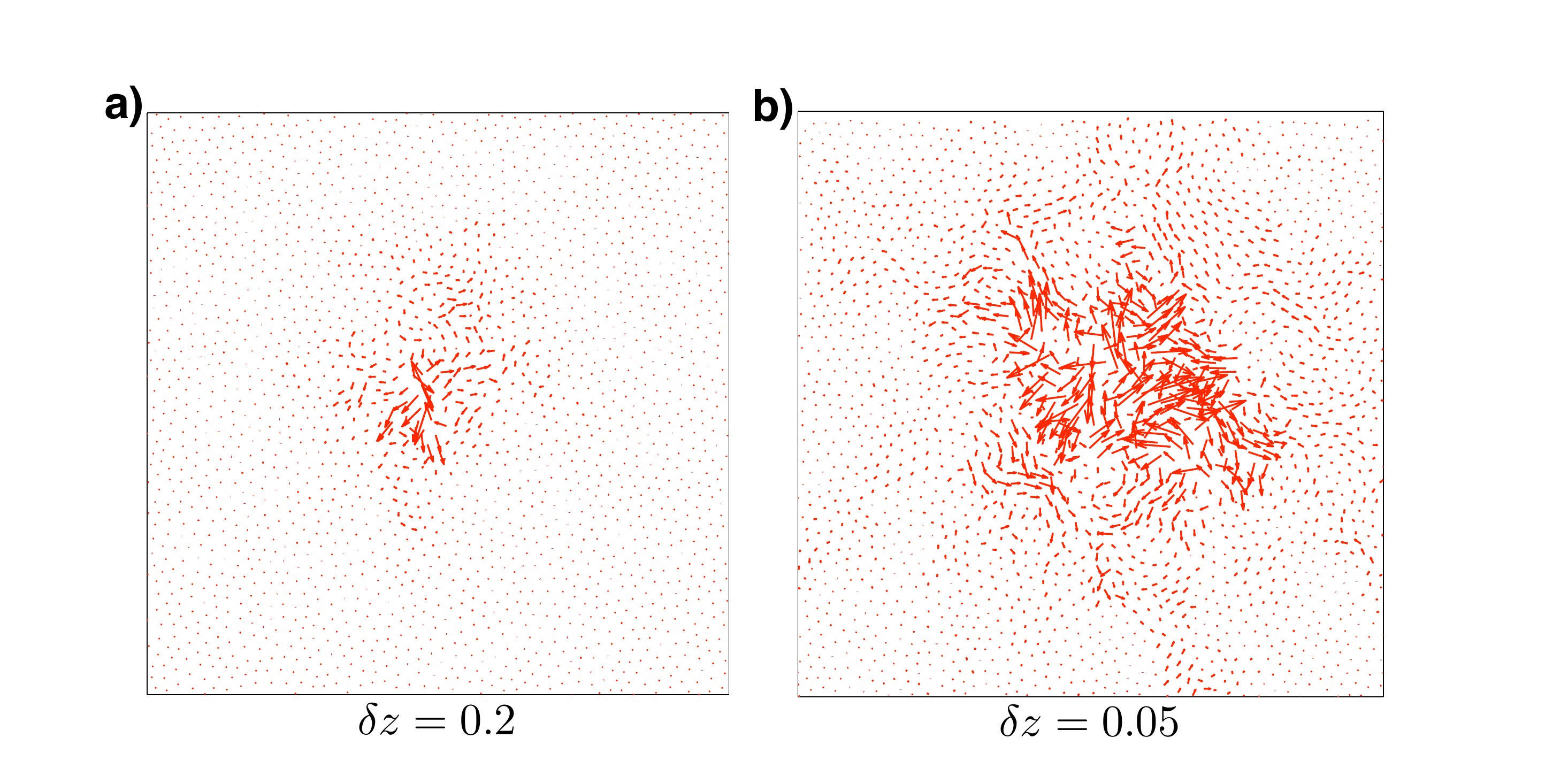} }
 \caption{  \label{figNonAffine} {Displacement field caused by  a local strain in a floppy network.  An over-damped relaxation was used following the elongation of one spring for (a) $\delta z=0.2$ and (b) $\delta z=0.05$. }}
 \end{center}
 \end{figure}
EMT attempts to describe  disordered materials as ordered materials with a frequency-dependent effective stiffness  ${\tilde k}_{\rm eff}$.
 We apply this technique to isotropic lattices of   coordination $z_{in}>>z_c$, where bonds are then  randomly removed with a probability $(1-p)$, so that the stiffness coefficients $k_{ ij}$ take the values $0$ or $k$ with a probability $(1-p)$ or $p$ respectively, and the final coordination is  $z=p z_{in}$.   The Green's function ${\bf G}(\omega)$ of the disordered floppy system is defined as $[-m\omega^2 +\mathcal{M}]{\bf G}(\omega)=-1$.
${\bf G}_0$  is the Green's function of the effective medium, corresponding to the initial ordered lattice with an undetermined  effective stiffness coefficient ${\tilde k}_{\rm eff}$. 
 Standard  calculations lead to the Dyson relation ${\bf G}={\bf G}_0+{\bf G}_0\mathcal{T}{\bf G}_0$,\cite{Webman} where
the operator $\mathcal{T}$ can be expressed as an infinite series in terms of increasing numbers of interacting contacts:
$$\mathcal{T}=\sum_{\langle ij\rangle} {\bf T}_{\langle ij\rangle}\quad+\sum_{\langle ij\rangle\neq\langle kl\rangle} {\bf T}_{\langle ij\rangle}{\bf G}_0{\bf T}_{\langle kl\rangle}+\ldots.$$
The transfer matrix is found to be:
\begin{equation}
{\bf T}_{\langle ij\rangle}=\frac{(\vert i\rangle-\vert j\rangle)(k_{ij}-{\tilde k}_{\rm eff})(\langle i\vert-\langle j\vert)}{1-(k_{ij}-{\tilde k}_{\rm eff}){\bf n}_{ij}(\langle i\vert-\langle j\vert){\bf G}_0(\vert i\rangle-\vert j\rangle){\bf n}_{ij}}{\bf n}_{ij} \otimes {\bf n}_{ij}. 
\label{T}
\end{equation}   

 We seek an effective stiffness ${\tilde k}_{\rm eff}$  that captures  the average behavior of the  system, i.e.  $\langle {\bf G}\rangle={\bf G}_0$, where the average is taken over the disorder  on the stiffness coefficients $k_{ ij}$.
This condition leads to $\langle\mathcal{T}\rangle=0$. 
In the EMT  this constraint is approximated by  $\langle {\bf T}_{\langle ij\rangle}\rangle=0$. 
Using standard identities for the Green's function on isotropic lattices,\cite{Kirkpatrick} one can express  this condition as:\cite{thorpe2}
\begin{equation}
m\omega^2 \mathrm{tr}[{\bf G}_0(r=0,\omega)]=\frac{\delta z+(z_{in}-2d)k_{\rm eff}}{2(1-k_{\rm eff})},
\label{effeqt}
\end{equation}
where  $\mathrm{tr}[\bullet]$ stands for the  trace, and  $k_{\rm eff}\equiv {\tilde k}_{\rm eff}/k$. 

Since we are interested in low frequencies $\omega<<\sqrt{k/m}$, we approximate ${\bf G}_0$ by its continuum limit and use a  Debye cut-off  $q_D$.  We introduce the bulk  
 modulus $K$ and the shear modulus $\mu$ of the ordered lattice without an effective stiffness, and use the approximation:
 
\begin{equation}
{\bf G}_0({\bf r},\omega)=\sum_\alpha \int_0^{q_D}d{\bf q}\frac{e^{i {\bf q}\cdot {\bf r}}}{m\omega^2-c_\alpha k_{\rm eff}  q^2} {\bf q}_\alpha \otimes {\bf q}_\alpha.
\label{GreenConti}
\end{equation}
The sum is taken over all polarization vectors $ {\bf q}_\alpha$, where $c_p=\frac{V}{N}(K+\frac{4}{3}\mu)$  for pressure waves and $c_s=\frac{V}{N}\mu$
for shear waves.

\subsection{Scaling analysis near $z=z_c$}
We now perform a scaling analysis of Eqs.(\ref{effeqt},\ref{GreenConti}) as $z\rightarrow z_c$ from below.
Eq.(\ref{GreenConti}) implies that $\mathrm{tr}[{\bf G}_0(0,\omega)]=\frac{1}{-k_{\rm eff}} f\left(\frac{m\omega^2}{-k_{\rm eff}}\right)$,  where the function $f$  is independent of $z$ and $\omega$.
As $\omega\rightarrow 0$, the elastic moduli of floppy systems   vanishes and   $k_{\rm eff}\rightarrow0$. Eq.(\ref{effeqt}) then  leads to
$\lim_{\omega\rightarrow0} \frac{m\omega^2}{-k_{\rm eff}}=\epsilon$, where $\epsilon$ satisfies the equation
 \begin{equation}
 \epsilon f(\epsilon)=\delta z/2.
 \label{epsilon}
 \end{equation}
For spatial dimensions  $d\geq3$, Eq.(\ref{GreenConti}) implies that $f(0)$ is a positive constant, therefore   in the regime  $\delta z<<1$ one finds $$\epsilon\approx \delta z/(2f(0)).$$  Using Eq.(\ref{effeqt}) together with the assumption that $|m\omega^2/k_{\rm eff}|<<1$ and $|k_{\rm eff}|<<1$ (which can be shown to be true {\it a posteriori} in the limit of  $\delta z<<1$ and $\omega<<\sqrt{k/m}$\,) one finds:
\begin{equation}
k_{\rm eff}\approx\frac{-\delta z+\sqrt{-8f(0)(z_{in}-2d)m\omega^2+\delta z^2}}{2(z_{in}-2d)}.
\label{keffR}
\end{equation}

For $d=2$, Eq.(\ref{GreenConti}) leads to  a logarithmic divergence in the small  $\epsilon$ limit:
\begin{equation}
f(\epsilon)\approx -\frac{V}{2N}(c_p^{-1}+c_s^{-1})\log(\epsilon)+\Gamma,
\label{ff}
\end{equation}
where $\Gamma$ is a constant that depends on the elastic moduli and the Debye cut-off.  The weak divergence
of $f$ does not modify the perturbation analysis performed for $d\geq 3$, 
except that $f(0)$ must now be replaced by $f(\epsilon)$ in Eq.(\ref{keffR}). Only  $f(\epsilon)$ remains undetermined.

The asymptotic value for $\epsilon$ in the limit $\delta z\rightarrow 0$  can be obtained  from Eq.(\ref{epsilon}), which leads 
to   $\epsilon\sim\frac{-\delta z}{\log(\delta z)}$ and therefore $f(\epsilon)\sim -\log(\delta z)$. This  
behavior is only  valid for values of $\vert \log(\delta z)\vert \gg1$, a difficult limit  to observe empirically.
Therefore, to compare our theoretical predictions with two dimensional numerical observations we compute $\epsilon(\delta z)$ by solving Eq.(\ref{epsilon}) numerically using 
Eq.(\ref{ff}). We find that the  relation $\epsilon(\delta z)$ depends on the initial network via only one parameter $\Lambda(c_p,c_s,\Gamma)$.\footnote{The parameter $\Lambda(c_p,c_s,\Gamma)=\log[\frac{V}{N}(c_p^{-1}+c_s^{-1})]+\frac{N\Gamma}{V(c_p^{-1}+c_s^{-1})}$} To compare the theory and numerical simulations, we use $\Lambda(c_p,c_s,\Gamma)$ as a fitting parameter.

The mechanical response of floppy systems is therefore fully  determined by the Green's function in Eq.(\ref{GreenConti}) and the effective
stiffness given by Eq.(\ref{keffR}). We start by computing the  density of states $D(\omega)$ using the relation 
$D(\omega)=\frac{2m\omega}{\pi}\mathrm{Im}[\mathrm{tr}[{\bf G}_0(0,\omega)]]$.  
For $0<\omega<\omega^*$, where 
$\omega^*=\frac{\delta z}{\sqrt{8f(\epsilon)(z_{in}-2d)m}}$,  we find that $D(\omega)=0$. For $\omega>\omega^*$ we obtain:
\begin{equation}
D(\omega)\approx\sqrt{\frac{2f(\epsilon)(z_{in}-2d)m}{\pi^2}}\sqrt{1-\left(\frac{\omega^*}{\omega}\right)^2} .
\label{density}
 \end{equation}  
Thus the size of the gap scales linearly with $\delta z$ for $d\geq3$,
where $f(\epsilon)\approx f(0)$, whereas for $d=2$,  a logarithmic correction exists.  
According to Eq.(\ref{density}), rescaling the frequency by $\omega^*$ and $D(\omega)$ by $f(\epsilon)^{1/2}$ should collapse the low-frequency part of the spectrum.
 Fig.\ref{DOSFI} shows that the quality of the  collapse is very good. For all considered coordinations, we used the same fitting parameter  $\Lambda(c_p,c_s,\Gamma)=1.3$, which is  fixed by this measurement.

  \begin{figure}[htb]
 \begin{center}
  \centerline{ \includegraphics[width=0.48\textwidth,trim=0mm 0mm 0mm 0mm, clip]{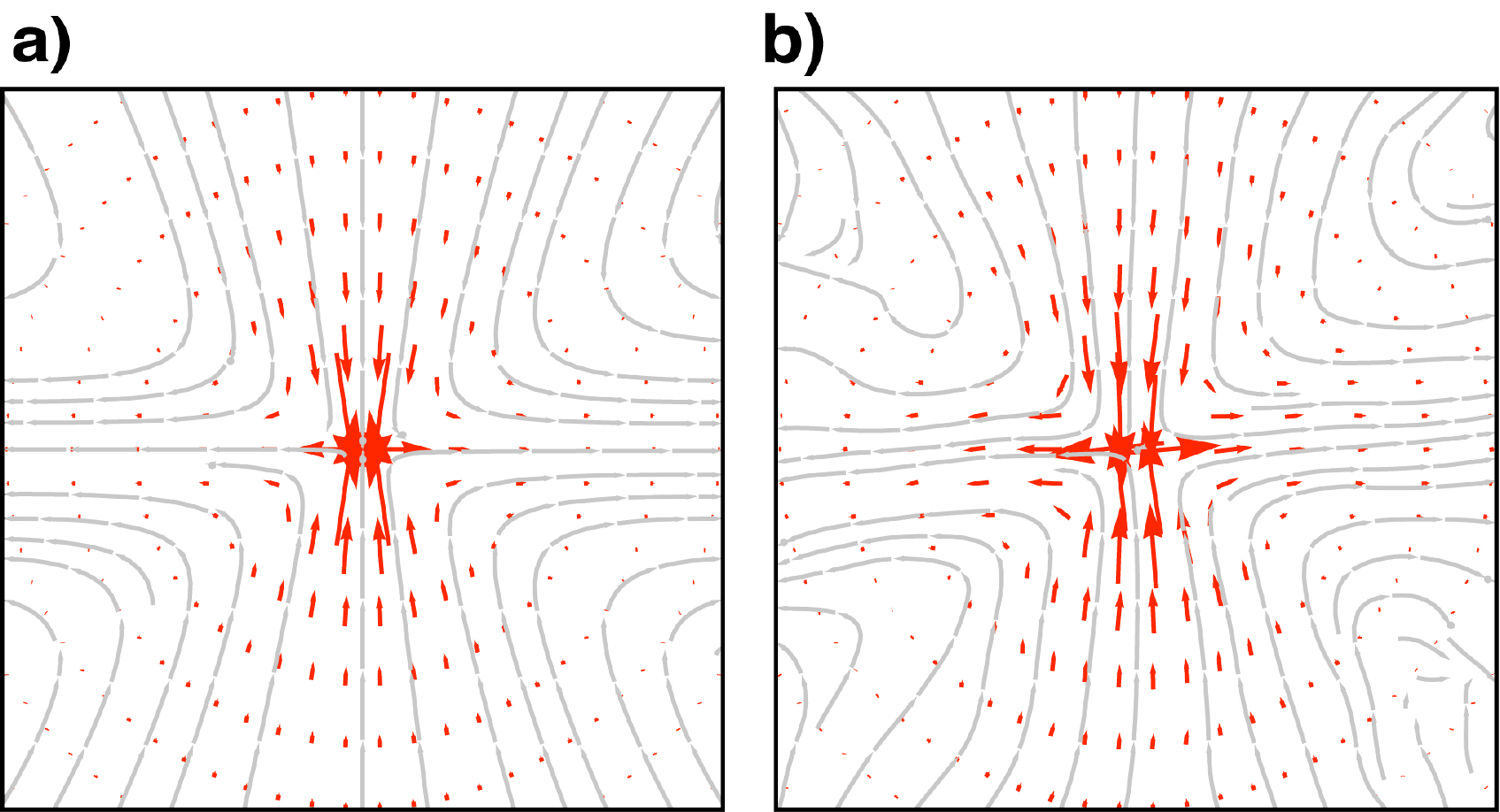} }
 \caption{  \label{mean} { Mean response following the elongation of a spring placed at the center, along the vertical axis. {\bf a)} Theoretical prediction given by Eq.~(\ref{decaymean})  using $c_s$  and $c_p$ of an hexagonal lattice. Arrows show the displacement field and continuous lines are  stream lines of the  vector field. {\bf b)} Average response over $6000$ independent realizations  of networks with $N=40000$ nodes and $\delta z=0.05$. }}
 \end{center}
 \end{figure}

\subsection{Linear response to a local strain}
 \label{LinearSub}
The response to a local force ${\bf G}_0(\omega,{\bf r}_{ij})$ diverges as $\omega\rightarrow 0$ due to the presence of floppy modes. 
One must rather consider the response field due to an imposed displacement. Changing the rest length of a spring placed between nodes $i$ and $k$ at a fixed frequency $\omega_0$ corresponds to imposing a displacement $(\delta {\bf R}_i-\delta {\bf R}_k)\cdot {\bf n}=e^{i\omega_0t}$, with ${\bf n}$ being the unit vector along the connecting bond. The force required to impose such a displacement is $$\vert F\rangle=\frac{ e^{i\omega_0 t}{\bf n}}{2{\bf n}\cdot({\bf G}_0(\omega_0,0) -{\bf G}_0(\omega_0,{\bf r}_{ki}))\cdot{\bf n}}(\vert k\rangle-\vert i\rangle). $$ 
In the small frequency regime, the last expression is found to be  $\vert F\rangle\sim k_{\rm eff}e^{i\omega_0 t}{\bf n}(\vert i\rangle-\vert k\rangle)$. The  magnitude of the force vanishes as $\omega_0\rightarrow 0$, which is consistent with the existence of floppy modes.

 The response  at the zero frequency limit   in two dimensions,  at distances $r=\vert{\bf r}\vert$ larger than the typical springs length (i.e. $r q_D\gg1$),  can be calculated using the continuum limit
 \begin{equation}
 \langle\delta {\bf R}({\bf r})\rangle={\bf g({\bf r}-{\bf r}_i)}-{\bf g({\bf r}-{\bf r}_k)},
  \label{decaymean}
\end{equation}
where  
$${\bf g({\bf r})}=\frac{1}{\pi\epsilon}\left[\nabla\times(\nabla\times K_0(r\sqrt{\frac{\epsilon}{c_s}}){\bf n})-\nabla(\nabla\cdot K_0(r\sqrt{\frac{\epsilon}{c_p}}){\bf n})\right]$$
and  $K_0$ is the modified Bessel function of the second kind, that behaves exponentially at long distances. Thus, the mean response  to a  local strain decays exponentially in floppy systems with a characteristic length $l_c\sim 1/\sqrt{\epsilon}$. The mean perturbation induced by changing the rest length of a spring has a quadrupolar symmetry shown in Fig.\ref{mean}a.   After averaging over 6000 realizations, we obtain  good agreement with our theoretical prediction, shown in in Fig. \ref{mean}b.

\begin{figure}[htb]
 \begin{center}
 \centerline{ \includegraphics[width=0.5\textwidth]{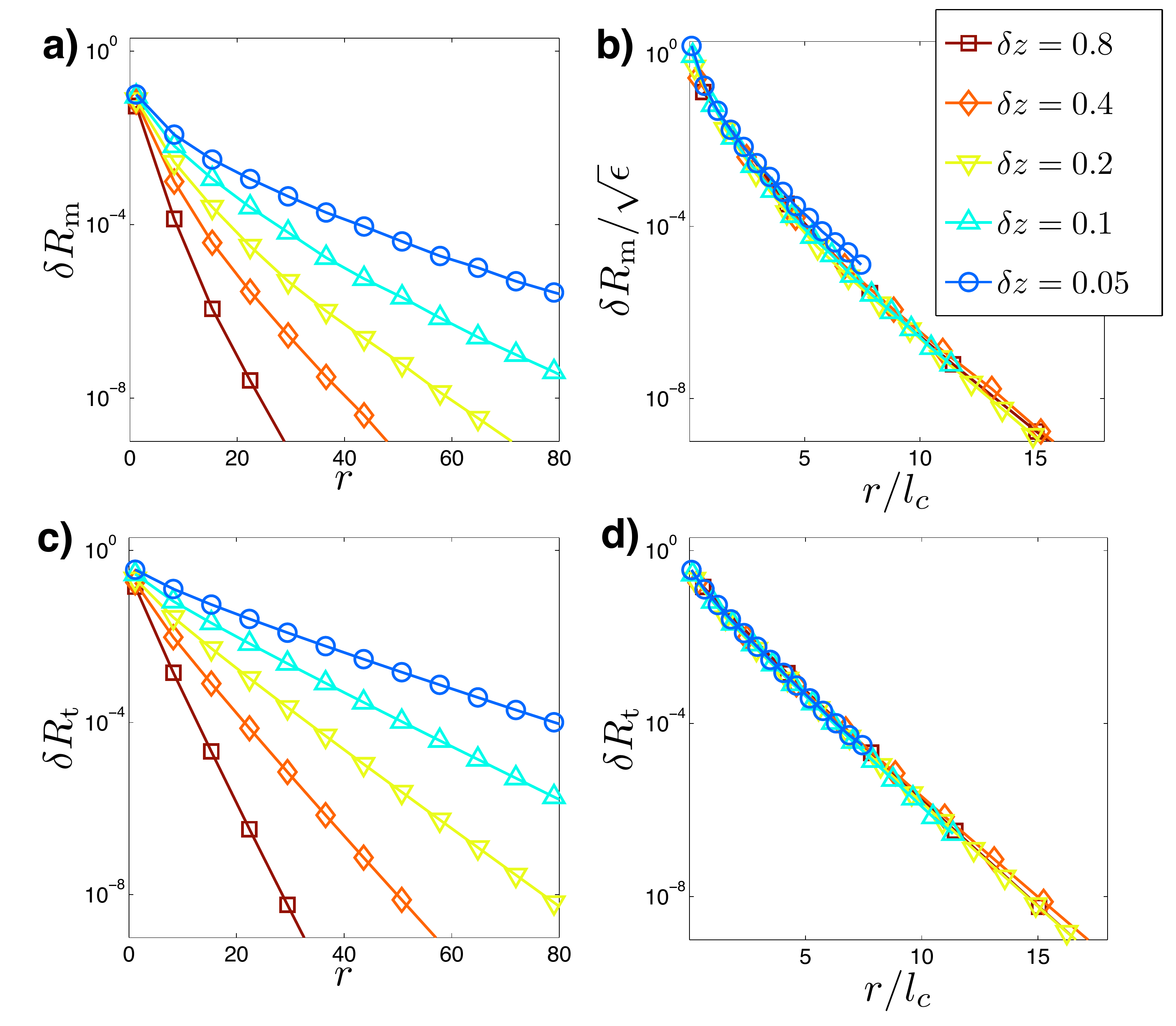} }
 \caption{  \label{compa} {{\bf a)} Mean response $\delta R_{\rm m}$ {\it vs.} the distance  $r$ from the imposed strain with  $N=40000$ nodes. {\bf b)} Rescaled Average displacement of the mean response  $\delta R_{\rm m}$ {\it vs.}  the rescaled distance $r/l_c$, using $\Lambda(c_p,c_s,\Gamma)=1.3$ as extracted from Fig.(\ref{DOSFI}).  {\bf c)} The fluctuations around the mean response are characterized by $\delta R_{\rm t}$, defined in Eq.(\ref{111}). $\delta R_{\rm t}$ is plotted {\it vs.} the distance from a local strain  $r$. {\bf d)} $\delta R_{\rm t}$ {\it vs.} the rescaled distance $r/l_c$, using $\Lambda(c_p,c_s,\Gamma)=1.3$. }}
 \end{center}
 \end{figure}

  The  asymptotic solution in any dimension has an exponential decay. Indeed, taking the angular average one gets 
  \begin{equation}
 \delta R_{\rm m}(r)\equiv  \overline{\langle\delta  {\bf R}(r)\rangle^2}^{1/2}\sim \epsilon^{(d-1)/2}e^{-r/l_c}(l_c/r)^{(d-1)/2},
   \label{meanAverage}
   \end{equation}
 where the over-line stands for angular average.  In two dimensional  networks, rescaling the distance by $l_c$ and the amplitude by $\sqrt{\epsilon}$ collapses  the response to a local strain at different coordination values into a single curve, as shown in Fig.\ref{compa}a-b.

The zero frequency limit  can be  extended to  finite frequencies $\omega<<\sqrt{k/m}$ by replacing $\epsilon \rightarrow -m\omega^2/k_{\rm eff}$ in equation (\ref{decaymean}). The asymptotic behavior is then given by  $\log( \langle\delta {\bf R}({\bf r})\rangle)\sim -r/l_c(\omega)+i\omega r /v(\omega)$, where 
$$l_c(\omega)=\frac{\vert k_{\rm eff}\vert}{\sqrt{m}\omega\vert \mbox{\bf Im}(k_{\rm eff}^{1/2})\vert}\qquad v(\omega)=\frac{\vert k_{\rm eff}\vert}{\sqrt{m}\mbox{\bf Re}(k_{\rm eff}^{1/2})}.$$ 
From Eq.(\ref{keffR}) one can prove the existence of two regimes. (i) For $\omega<\omega^*$,  the imaginary part  $1/v(\omega)=0$: one finds a pure exponential decay, with a characteristic length of order $ l_c$.  (ii) For $\omega>>\omega^*$, the  decay length decreases as $l_c(\omega)\sim f(\epsilon)^{1/4}\omega^{-1/2}$, whereas the velocity of the corresponding vibrations grows as $v(\omega)\sim f(\epsilon)^{1/4} \omega^{1/2}$. A similar behavior above $\omega^*$  has been  predicted by one of us for $z\geq z_c$.\cite{wyart2010} 

 We can also considering the case of our networks immerse in a viscous fluid with viscosity $\eta_0$ and without hydrodynamic interactions. The response in the over-damped regime is  calculated by replacing the inertial term $m\omega^2$ by the viscous force $-i\eta_0\omega$ in equation (\ref{GreenConti}) and then in the effective stiffness (\ref{keffR}). In this limit the elastic moduli is proportional to $k_{\rm eff}$, whose real and imaginary parts gives the storage and loss moduli $G'$ and $G''$ respectively.  Two different regimes are observed. (i) For $\omega<<{\omega^*}^2m/\eta_0$, the storage moduli follows  $G'\sim l_c^6\omega^2/f(\epsilon)$ and the loss moduli $G''\sim l_c^2\omega$. The response is given by   $l_c( \omega) \sim l_c$ and $v(\omega)\sim f(\epsilon) l^{-3}_c$,   displaying a similar exponential decay as the one observed in undamped system, but with propagating waves. (ii) For  $\omega>>{\omega^*}^2m/\eta_0$ both, the storage and loss moduli are proportional to $\sqrt{f(\epsilon)\omega}$, as also  observed in Ref.\cite{brian2}  The decaying length    $l_c(\omega)\sim f(\epsilon)^{1/4}\omega^{-1/4}$, while the velocity  $v(\omega)\sim f(\epsilon)^{1/4} \omega^{3/4}$.  Our scaling for the characteristic length scale $l_c(\omega)$ differs from the recent result of Tighe.\cite{brian2} In our opinion, the difference stems from an incorrect definition of the length scale characterizing the elastic response in floppy materials  (see  footnote\footnote{ Tighe\cite{brian2}  defines a  length scale $\lambda_f=f/F$ as the ratio between the typical contact force $f$ carried by springs to the viscous forces $F$ exerted by the fluid. $F$  is proportional to the velocity of the particles, and therefore to their displacements time the frequency $\omega$. It is thus readily extractable from our results. The contact forces $f$ are such that forces are balanced on each node, which implies {\it on average} that the spatial derivative of the contact forces $f$ are the external forces $F$. If one considers for example the response to a local perturbation, our result implies that $\langle f\rangle/\vert \langle {\bf F}\rangle\vert$ is of order of the length scale $l_c(\omega)$ on which the mean elastic response decays. Thus this definition is  consistent with our results. However a  different result is obtained if one considers the ratio $\sqrt{\langle f^2\rangle}/\sqrt{\langle F^2\rangle}$, as done by Tighe. These two quantities differ because the mean response is much smaller than the fluctuations around it (see  section \ref{fluctuation}). Therefore, associating the quantity  $\lambda_f=\sqrt{\langle f^2\rangle}/\sqrt{\langle F^2\rangle}$ to a length appears unjustified. In the zero frequency limit we can actually calculate this last expression using the formalism developed in \cite{edan}. For isotropic random floppy networks one gets $\langle f^2\rangle\sim {\omega^*}^{-3}$ and $\langle F^2\rangle\sim {\omega^*}^{-1}$, leading to a ratio $\sqrt{\langle f^2\rangle/\langle F^2\rangle}\sim{\omega^*}^{-1}\sim l_c^2$ consistent with the numerical results of Tighe.}).

\section{Beyond mean field}
 \label{BMF}

\subsection{Fluctuations}
\label{fluctuation}
The obvious difference between the mean response to a local strain (Fig.\ref{mean}) and a typical one (Fig.\ref{figNonAffine}) indicates large fluctuations.   A priori mean field models as the effective medium do not enable to capture those. However, it is possible to combine  EM results   with  additional considerations to estimate the amplitude of these fluctuations.

We denote  the dipole of forces  $\vert  {\bf F}_{ij}\rangle$ generated by changing the rest length of the spring $ij$ by a distance one:
\begin{equation}
\vert  {\bf F}_{ij}\rangle={\bf n}_{ij} (\vert i\rangle-\vert j \rangle),
\label{Fdipole}
\end{equation}
where the stiffness coefficient $k$ is set to unity.   
From the definition of the stiffness matrix, Eq.(\ref{00}),  and Eq.(\ref{Fdipole}), we can write:
\begin{equation}
\label{33}
 \mathcal{M}= \sum_{\langle ij\rangle}\vert {\bf F}_{ij}\rangle \langle{\bf F}_{ij}\vert,
 \end{equation}
 where the sum is taken over all the bonds.  Note that any floppy mode $\vert \delta{\bf R}_0\rangle$ has by definition no restoring force $\mathcal{M} \vert \delta{\bf R}_0\rangle=0$, implying that $\langle\delta{\bf R}_0 \vert  {\bf F}_{ij}\rangle=0$ for all contacts $ij$.

Thus, the response $\delta{\bf R}_{ij}$ to the elongation of a spring $ij$, which is equivalent to the response to a force dipole $\vert  {\bf F}_{ij}\rangle$, has no components along floppy modes.  Therefore the equation ${\cal M} \delta{\bf R}_{ij}=  \vert  {\bf F}_{ij}\rangle$ can be inverted. Using the spectral decomposition of ${\cal M}$, one gets:
\begin{equation}
\label{22}
\vert \delta{\bf R}_{ij}\rangle=\sum_{\omega>0}\frac{1}{ \omega^{2}}\vert\delta{\bf R}_\omega\rangle\langle\delta{\bf R}_\omega\vert {\bf F}_{ij}\rangle,
\end{equation}
 where   $\omega^2$ and $\vert\delta{\bf R}_\omega\rangle$ are the non-zero eigenvalues and the corresponding  eigenvectors  of the stiffness matrix ${\mathcal M}$.\cite{edan,Wyartmaha}   Eq.(\ref{22}) implies that the norm of the response follows 
$\langle \delta {\bf R}_{ij}\vert \delta {\bf R}_{ij} \rangle=\sum_{\omega>0} \langle\delta{\bf R}_\omega\vert {\bf F}_{ij}\rangle^2/\omega^4$. 
Introducing the average amplitude of the response  $\delta R^2_d\equiv\frac{1}{N_c}\sum_{\langle ij\rangle} \langle \delta {\bf R}_{ij}\vert \delta {\bf R}_{ij} \rangle$, one gets using Eq.(\ref{33}):
\begin{eqnarray}
\delta R^2_d&=&  \frac{1}{N_c}\sum_{\omega>0} \sum_{\langle ij\rangle}  \frac{\langle\delta{\bf R}_\omega\vert {\bf F}_{ij}\rangle^2}{\omega^4} =    \frac{1}{N_c}\sum_{\omega>0} \frac{1}{\omega^4}   \langle\delta{\bf R}_\omega\vert{\cal M}|\delta{\bf R}_\omega\rangle\notag\\
&=& \frac{1}{N_c}\sum_{\omega>0}\frac{1}{\omega^2}\sim\int_{\omega^*} \frac{ D(\omega) }{ \omega^2}d\omega\sim\frac{1}{\epsilon}.
\label{TDis}
\end{eqnarray}
For comparison, the total amplitude of the mean  response can be  calculated from Eq.(\ref{meanAverage}), and one obtains for the norm square $\int\delta R_m(r)^2d^dr\sim \epsilon^{\frac{d}{2}-1}$. Thus for $d\geq 2$, the norm of the mean response vanishes as $\delta z\rightarrow0$, whereas the norm of the fluctuations diverge. Thus relative fluctuations must diverge at small $\epsilon$, as observed in our data.  

The most simple scenario is that the fluctuations of the response $\langle \delta { R}({\bf r})^2\rangle$ decays  with the same characteristic length $\propto l_c$ characterizing the mean response.  Making this assumption, which is numerically verified (see below), and using the result of Eq.(\ref{TDis}) , the  angular average must read:
\begin{equation}
\label{111}\delta R_{\rm t}(r)\equiv \overline{\langle\delta  R(r)^2\rangle}^{1/2}\sim \epsilon^{(d-2)/4}h(r/l_c),
\end{equation}
 where  $\log(h(x))\sim -x$ for $x>>1$, and the $\epsilon$ dependence is determined by Eq.(\ref{TDis}).
  Rescaling the distance by $l_c$  leads to a very good collapse of the response  at different coordination values, as shown in Fig. (\ref{compa}c-d). Note that in two dimensions the amplitude does not need to be rescaled.

  \begin{figure}[htb]
 \begin{center}
  \centerline{ \includegraphics[width=0.48\textwidth,trim=0mm 5mm 0mm 0mm, clip]{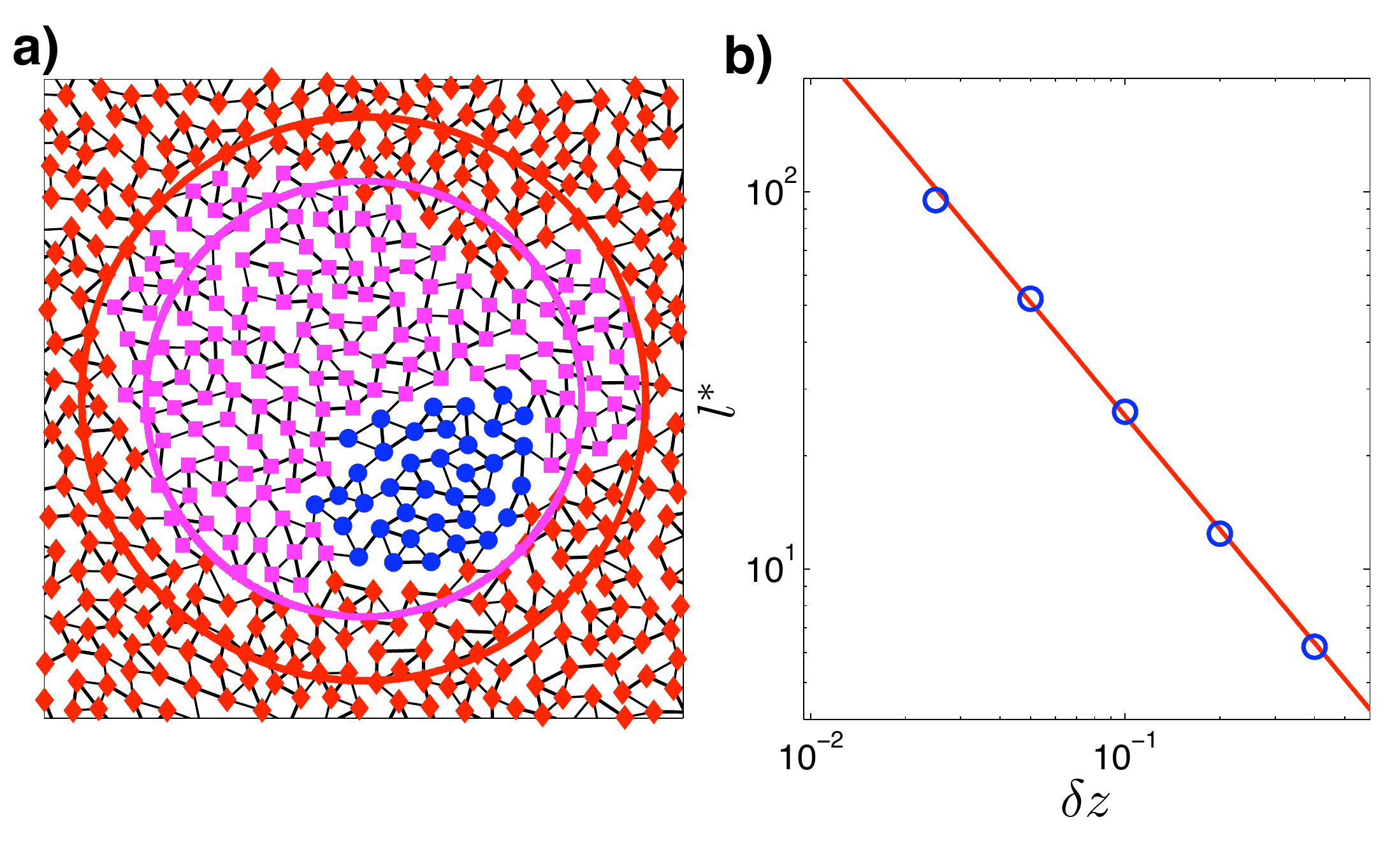} }
 \caption{  \label{pebble} { Rigid regions, shown as red diamonds and magenta  square,  are induced  by freezing the nodes outside an external  radius (red)   and an internal radius (magenta)  respectively. Circular nodes (blue)  show the minimal floppy region. Freezing any extra particle rigidifies the entire system.  (b) $l^*$, as defined in the text, {\it vs.} $\delta z$. Red line corresponds to  $l^*\sim 1/\delta z$.
 }}
 \end{center}
 \end{figure}
 \subsection{Pinning boundaries}
We now turn to the spatial properties of floppy modes. The response to the stretch of a spring, exemplified in  Fig.(\ref{figNonAffine}), is a floppy mode of the network where this spring is removed.  
We have thus shown that floppy modes can be localized (in the sense of presenting an exponential decay) on a length scale $l_c\sim 1/\sqrt{ \delta z}$  (with a log correction for $d=2$).
We now extend  a previous counting argument \cite{Wyart1,Wyart053}, tested for packings with $z>z_c$ in \cite{BULBUL}, to  floppy networks (see also \cite{brian2}) and show that floppy modes are also characterized by another length scale  $l^*\sim 1/\delta z>>l_c$. Below $l^*$  {\it strict} localization is impossible: floppy modes cannot have a smaller compact support.
Essentially, $l^*$ is the length at which the number of constraints $r_f^{d-1}$  that result  from freezing the boundary of a system of size $r_f$  is
equal to the number of floppy modes in the bulk $\delta z \, r_f^d$.  This counting argument suggests that if the boundaries are frozen on a scale $l^*$ or smaller, floppy modes
must vanish. 

To test this prediction, we fix all the  nodes outside a circle of radius $r_f$. A frozen boundary induces 
a rigid region where floppy modes are forbidden. We determine this region using the pebble 
algorithm,\cite{pebble} as shown in Fig.(\ref{pebble}.a).  As $r_f$ decreases, all floppy modes eventually vanish. For any $z$ we can define $n^*$, the number of nodes involved in the last floppy mode, and $l^*\equiv n^*{}^{1/d}=\sqrt{ n^*}$ for $d=2$. Our measurements are shown in  Fig.(\ref{pebble}.b) and follow the prediction $l^*\sim 1/\delta z$. This  result implies that exponentially small displacement at distances  $r>l_c$  cannot be neglected  when the rigid-floppy transition induced by freezing boundary conditions is considered. On the other hand, our results imply that floppy networks that are stabilized by pinning boundaries on the scale $l^*$ present soft modes with exponentially small frequencies near the center of the sub-system, since there are floppy modes with displacements of very tiny amplitude near the boundaries, of order exp$(-l^*/l_c)\sim$ exp$(-1/\sqrt\delta z)$.

\begin{figure}[htb]
 \begin{center}
 \centerline{ \includegraphics[width=0.53\textwidth,trim=0mm 0mm 0mm 0mm, clip]{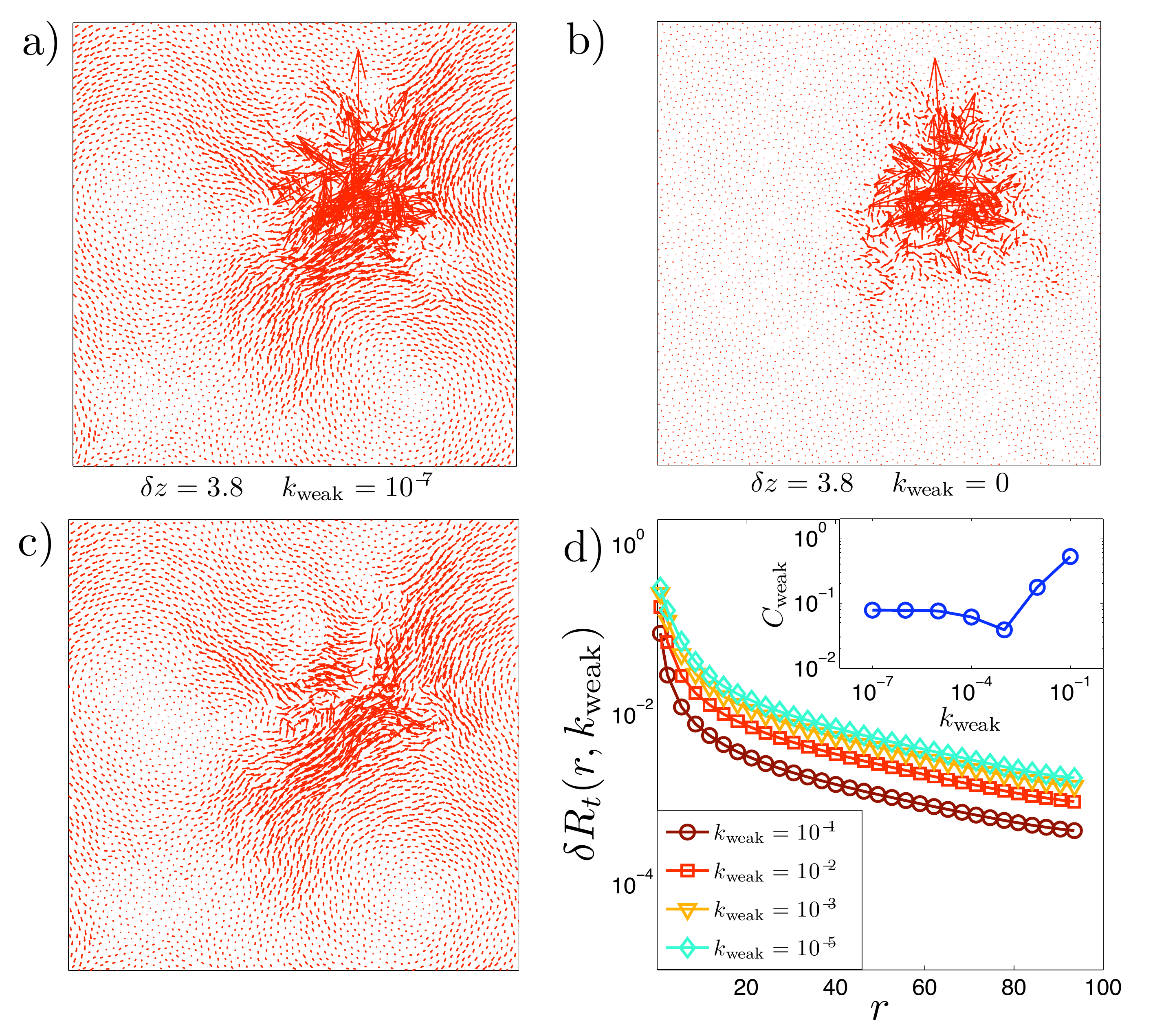} }
 \caption{  \label{weakA} { {\bf a)} Example of the response to a local strain for $\delta z =0.2$ and $k_{WEAK}=10^{-7}$. {\bf b)} response to the same strain in the absence of weak spring ($k_{weak}=0$), as studied  in the last sections. {\bf c)} Difference between the displacement field of  {\bf a)} - {\bf b)}.  {\bf d)} Average displacement $\delta R_{\rm t}$  {\it vs.} the distance $R$ from a local strain   for  floppy networks ($\delta z=0.2$) with weak springs, as indicated in legend, with $N=90000$ nodes. Inset: $C_{\rm weak}$  {\it vs.} $k_{\rm weak}$  (see text for definition),  where $L_0$ has been taken to be $80$ spring lengths. }}
 \end{center}
 \end{figure}

\section{Weak interactions}
\label{WIF}

As discussed in introduction,  in some materials floppy modes  are stabilized by weak interaction, as is the case in covalent glasses. In order  to model these weak interactions,  we consider  (see for example \cite{Wyartmaha})  floppy networks  with $k=1$ and add a  number of weak springs of  stiffness $k_{\rm weak}\ll m{\omega^*}^2$, which gives a finite elasticity to the network. Floppy modes  then gain  finite frequencies, of order the characteristic frequency scale associated with the weak interaction $\omega_c\equiv \sqrt{k_{\rm weak}/m}$. A gap in the density of states remain present in the frequency range $[\omega_c, \omega^*]$, if the weak interaction is weak enough, i.e. $\sqrt{k_{\rm weak}}/m<<\omega^*$. In this limit the elastic moduli   scale as $\mu_{\rm weak}\sim k_{\rm weak}/\delta z$,\cite{Wyartmaha}  where $\delta z$ is the excess coordination of the network of strong interaction. At long enough wavelengths the material, to  good approximation, must behave as a continuous elastic medium, and for $\omega<<\omega_c$ the density of states must follow a Debye behavior $D(\omega)\sim \omega^{d-1}$. Extracting a velocity of sound from the elastic moduli and computing the wavelength at $\omega\sim \omega_c$, one obtains a wave length of order $l_c$, which thus characterizes as well the length scale above which a continuum description becomes a good approximation.  

\begin{figure}[htb]
 \begin{center}
 \centerline{ \includegraphics[width=0.48\textwidth,trim=0mm 0mm 0mm 0mm, clip]{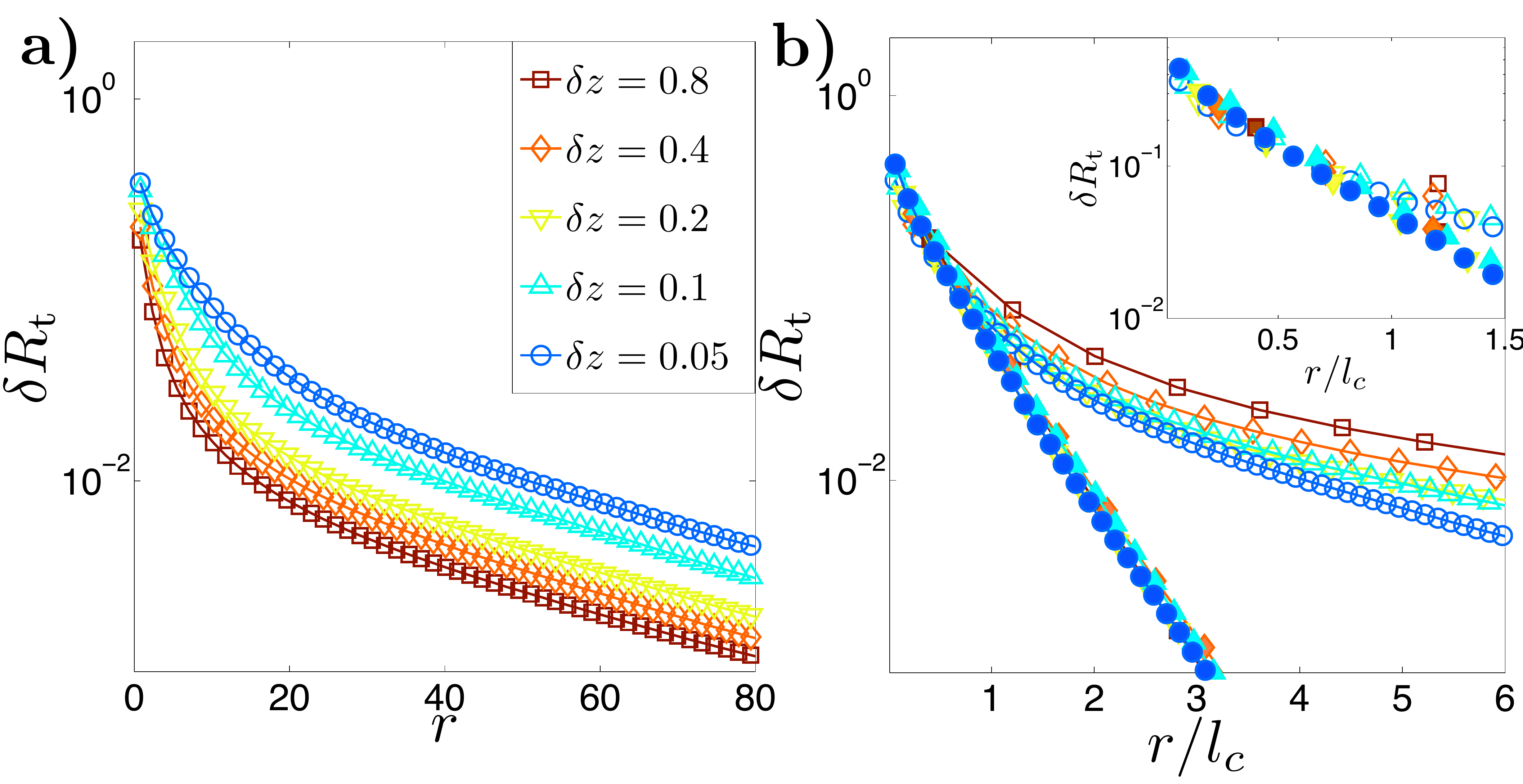} }
 \caption{  \label{weak} {{\bf a)} Average displacement $\delta R_{\rm t}$  {\it vs.} the distance from a local strain  $r$ for  floppy networks plus weak springs (see text) with $N=90000$ nodes and $k_{\rm weak}/k=10^{-7}$. {\bf b)} Average displacement $\delta R_{\rm t}$  {\it vs.}  the rescaled distance $r/l_c$, using $\Lambda(c_p,c_s,\Gamma)=1.3$. The response of strictly floppy networks (filled symbols) has been included for comparision. Inset: The initial decay magnified. }}
 \end{center}
 \end{figure}

 In Fig.\ref{weakA}a   the  response to a local strain for a system with weak springs with $k_{\rm weak}=10^{-7}$ is shown. The deformation field clearly differs from the response of the same network with $k_{\rm weak}=0$ (Fig.\ref{weakA}b).
We now argue that   the limit  $k_{\rm weak}\rightarrow0$ is singular: for any $k_{\rm weak}>0$, the response decays as a power-law at large distances $r>>l_c$,  even  in the limit $k_{\rm weak}\rightarrow 0$. The singularity of this limit can be seen by decomposing the response to a local strain in two parts. First, we consider the displacement of characteristic amplitude $\delta R$ that would follow such a strain in the absence of weak interactions, i.e. with $k_{\rm weak}=0$  (Fig.\ref{weakA}b). Second, we consider the additional displacement induced by the presence of weak springs  (Fig.\ref{weakA}c). Indeed due to the weak springs forces are not balanced after step one, and forces of order  $ F_{\rm weak}\sim k_{\rm weak} \delta  R $ have appeared on the nodes. In the limit $k_{\rm weak}\rightarrow 0$, these forces are vanishingly small and lead to no relaxation on the modes of non-vanishing  frequency $\omega>\omega^*$.  However the spectrum now presents modes (stemming from the floppy modes that exist as $k_{\rm weak}=0$) of characteristic frequency $\omega_c$. These modes relax due to the unbalanced weak forces with  amplitude    $x$ that must satisfy $m \omega_c^2 x\equiv k_{\rm weak}x\sim F_{\rm weak}\sim k_{\rm weak} \delta  R $,  implying that $x$ is independent of $k_{\rm weak}$.  The convergence of the response $\delta R_t(r,k_{\rm weak})$ in the small  $k_{\rm weak}$ limit is shown in  Fig.\ref{weakA}d. We can define $$C_{\rm weak}=\frac{\int_0^{L_0}\vert{\delta R_t(r,k_{\rm weak})-\delta R_t(r,0)\vert^2}dr}{\int_0^{L_0}\vert{\delta R_t(r,0)\vert}^2dr}$$ to quantify the difference between the response of the networks with weak springs and the ones strictly floppy. In the Inset of  Fig.\ref{weakA}d one can clearly see that $C_{\rm weak}$ converge to a non-zero value showing the singularity of this limit.

Although localization is lost, the coordination continues to influence the response to a local strain (see Fig.\ref{weak}a). In particular, near the imposed strain for $r<<l_c$, the response is weakly-influenced by the weak interaction, and decays with the same characteristic length, as appears in Fig.\ref{weak}b. Note that we expect this scenario to hold also for gently compressed, hypostatic soft ellipses\footnote[2]{In the case of ellipses the pre-stress \cite{alexander}  stabilizes  the system, shifting the zero frequencies of floppy modes  to   finite values\cite{cory2}. We believe that the pre-stress plays a similar role to the  weak springs of our network model. Accordingly,  we expect the response to a local strain to be extended, even when the pre-stress is vanishingly small. Finally, note that our theoretical description of the vibrational spectrum does not capture the singular coupling that occurs between translational and rotational modes when the ellipticity $\alpha$ is small. We thus expect our approach to apply only for rather large values of $\alpha$. }.

\section{Rheology of dense suspensions}
\label{RDS}

\subsection{ Random networks under a global shear }

It is interesting to consider the behavior of the present networks under shear, if they were placed in a viscous solvent. This problem is formally related  \cite{edan} to the affine solvent model of suspension flow of hard particles,
where hydrodynamic interactions between particles are neglected. We consider that our networks are made of point particles, subjected to a viscous drag ${\bf F} $  proportional to the particles velocities  ${\bf V}$ in the reference frame of the solvent, which is assumed to follow an affine shear. One can show, from the definition of this model,  that the viscosity $\eta$ is proportional to the non-affine velocity squared of the particles.\cite{edan, bruno,brian2}
The viscosity is proportional to the ratio $P/\dot{\gamma}^2$,  where $P$ is the total power dissipated  and $ \dot{\gamma}$ is the strain rate. The  power dissipated follows  $P\sim \langle {\bf F} \cdot{\bf V}\rangle\sim ||{\bf V}||^2$. It is convenient to write the velocity field in terms of the non-affine displacement following an infinitesimal strain $\delta {\bf R}/\delta \gamma$, i.e. ${\bf V}=\dot{\gamma} \delta {\bf R}/\delta \gamma$.  One find that $\eta \sim \langle (\delta R/\delta \gamma)^2 \rangle$, which relates the viscosity to the non-affine response to shear. 

For isotropic floppy  networks It was found numerically  \cite{Wyartmaha} that $||\delta {\bf R}/\delta \gamma||\sim 1/\sqrt \delta z$, implying $\eta\sim 1/\delta z$. Using our previous result on the response to a dipole in floppy materials and a simple hypothesis it is straightforward  to derive this result, thus extending a previous derivation valid for $z>z_c$,\cite{Wyartmaha} see also.\cite{brian2} Here we perform this calculation for completeness, and because the present argument can be extended to predict that the correlation length under shear is $l_c$. 
We consider  an affine shear strain $\delta \gamma$ applied on the network. After such a strain, unbalanced forces appear on the nodes:  $\vert  {\bf F}_{\delta\gamma}\rangle=\sum_{\langle ij\rangle }\gamma_{ij}\vert  {\bf F}_{ij}\rangle,$ where the sum is taken  over all the bonds and  $\vert  {\bf F}_{ij}\rangle$ correspond to a  dipole of force as defined in (\ref{Fdipole}). The coefficients $\gamma_{ij}$ are equal to ${\bf n}_{ij}\cdot\Gamma\cdot{\bf n}_{ij}$ where   $\Gamma$ is  the strain tensor, which is linear in $\delta \gamma$ at  first order.  The displacement field can be written as a linear combination of  responses to local dipoles   $\vert\delta{\bf R}_{\delta\gamma}\rangle=\sum_{\langle ij\rangle }\gamma_{ij} \vert \delta {\bf R}_{ij}\rangle$, where $\vert\delta {\bf R}_{ij}\rangle$ is defined in Eq.(\ref{22}). Two-point displacement correlations in space obey 
 \begin{multline}
C_{\delta\gamma}({\bf r})\equiv\langle \delta{\bf R}_{\delta\gamma}({\bf x}+{\bf r})\cdot \delta{\bf R}_{\delta\gamma}({\bf x})\rangle=\\
\sum_{\langle ij\rangle \langle kl\rangle} \langle\gamma_{ij}\gamma_{kl}  \delta{\bf R}_{ij}({\bf x}+{\bf r})\cdot \delta{\bf R}_{kl}({\bf x})\rangle.
\label{correla}
 \end{multline}
We now make the assumption that in random isotropic networks  (as those considered here), the response of different dipoles is  weakly-correlated (this assumption turns out to be incorrect for  flow of particles where subtle correlations are present in the structures visited by the dynamics). Using this assumption:  
 \begin{equation}
 C_{\delta\gamma}({\bf r})\sim\delta \gamma^2\sum_{\langle ij\rangle}\langle\delta{\bf R}_{ij}({\bf x}+{\bf r})\cdot \delta{\bf R}_{ij}({\bf x})\rangle, 
 \label{correla2}
 \end{equation}
 where we used $\gamma^2_{ij}\sim\delta\gamma^2$. 
Combining with  Eq.(\ref{TDis}) one finds
 $$   C_{\delta\gamma}(0)=\langle \delta{\bf R}_{\delta\gamma}({\bf x})\cdot \delta{\bf R}_{\delta\gamma}({\bf x})\rangle\sim\delta \gamma^2\delta R^2_d\sim \frac{\delta \gamma^2 }{\epsilon},$$
implying that the viscosity diverges as $\eta\sim C_{\delta\gamma}(0)/\delta \gamma^2\sim\epsilon^{-1}$. 
 
Moreover, (\ref{correla2}) indicates that the correlation length in (\ref{correla2}) is essentially the length scale appearing in the response to a dipole $l_c$.  $\langle\delta{\bf R}_{ij}({\bf x}+{\bf r})\cdot \delta{\bf R}_{ij}({\bf x})\rangle$ must vanish when $r$ is larger than the length $l_c$ where the response to a dipole is localized. On the other hand, correlations are not expected to vanish on a scale much smaller  than $l_c$, since the mean dipolar response will already give correlations on that scale: $$\frac{\delta \gamma^2}{N}\sum_{\langle ij\rangle}\int\langle\delta{\bf R}_{ij}({\bf x}+{\bf r})\rangle\cdot \langle\delta{\bf R}_{ij}({\bf x})\rangle d{\bf x}\sim e^{-\frac{\vert {\bf r}\vert}{l_c}}.$$

\subsection{ ASM model of suspension flows }

We predict a plateau of modes above a frequency $\omega^*\sim \delta z$. This plateau  also appears in models of suspension flows.\cite{edan} This observation supports that the plateau is generically present in floppy materials, and that the frequency response above $\omega^*$ will be well described by the effective medium predictions derived here.  There is, however, a crucial difference between the spectra of our networks and those of shear flows: in the latter one mode appears at a frequency $\omega_{\mathrm{min}}<<\omega^*$ that strongly couples to a global shear and leads to a divergence of the viscosity $\eta\sim 1/\delta z^{2.85}$,\cite{edan} which is  sharper than in isotropic networks where $\eta\sim 1/\delta z$. Therefore, typical configurations visited in dense flows differ qualitatively from our  random networks, for reasons explained in Ref.\cite{edan2} Nevertheless, we expect some of our present predictions to hold in flow.

For instance, the counting argument comparing surface effects and bulk degrees of freedom should hold,  and  $l^*$ is expected to characterize the effect of freezing boundaries. 

We do not know what the spatial correlations of the velocities should be in flows, because in this case the velocity is dominated by the lowest frequency mode that is not captured by the present analysis.\cite{edan,edan2} However the response  to a local disturbance (such as the formation of a new contact)   must couple predominantly to the modes above $\omega^*$,  and  is thus expected to affect the flow on a length $l_c$. This can be shown using  spectral decomposition for the response under a dipole of force $ \vert {\bf F}_{ij}\rangle$. Decomposing the spectrum in two parts  (above and below  $\omega^*$) Eq.(\ref{22}) reads
\begin{multline}
\vert \delta{\bf R}_{ij}\rangle=\sum_{\omega_{\mathrm{min}}<\omega<\omega^*}\frac{1}{ \omega^{2}}\vert\delta{\bf R}_{\omega}\rangle\langle\delta{\bf R}_{\omega}\vert {\bf F}_{ij}\rangle\\ +\sum_{\omega>\omega^*}\frac{1}{ \omega^{2}}\vert\delta{\bf R}_\omega\rangle\langle\delta{\bf R}_\omega\vert {\bf F}_{ij}\rangle.
\label{flow1}
\end{multline}
Numerical results show the appearance of a single mode below $\omega^*$,\cite{edan} however a theoretical estimation in the large $N$ limit supports that the distribution of modes between $\omega_{\mathrm{min}}$ and $\omega^*$ is given by $D_{\mathrm{min}}(\omega)\propto \omega(\omega^2-\omega_{min}^2)^{(d-2)/2}$.\cite{edan2}  Following the argument of section \ref{fluctuation},  the average amplitude of the response  is given by 
  $$\frac{1}{N_c}\sum_{\langle ij\rangle} \langle \delta {\bf R}_{ij}\vert \delta {\bf R}_{ij} \rangle\sim \underbrace{\int^{\omega^*}_{\omega_{\mathrm{min}}}\frac{D_{\mathrm{min}}(\omega)}{\omega^2}d\omega}_{I_{\mathrm{min}}}+\underbrace{\int_{\omega^*} \frac{D(\omega)}{\omega^2}d\omega}_{I^*}.$$
  
 The relative contribution follows $I_{\mathrm{min}}/I^*\sim \delta z^{(d-1)}$ (with a a logarithmic correction in two dimensions), indicating that in the limit $\delta z<<1$ the contribution of the modes below $\omega^*$ becomes vanishingly small and can be neglected. The modes above $\omega^*$ in flow have the same density of states and are expected to have the same properties than those of isotropic networks. Thus the response to a local disturbance in flow must decay exponentially with a typical length $l_c$, as could be tested empirically in two dimensional granular flows where imaging is possible, or perhaps using confocal imaging in slow emulsion flows. This simple argument does not hold for the velocity  correlation in flow: the modes in the plateau do not contribute significantly to the response to shear, which is rather dominated by the lowest frequency modes.

\section{ Discussion and open questions}

We have argued that the elasticity of floppy networks is characterized by a gap in the vibrational spectrum, and by two length scales $l_c$ and $l^*$ that diverge near jamming (i.e. $\delta z\rightarrow 0$). The existence of two lengths raises the question of which properties are governed by which scale. Our work supports that  the length scale characterizing the response to imposed forces, and to most standard observations, is $l_c$.  On the other hand $l^*$ characterizes the effect of pinning boundaries. Such effects are subtle, and can depend on surprising ways on the type of elastic networks considered.\cite{moukarzel3}.

 For the strongly disordered floppy isotropic networks that we consider, one example of a question that remains to be explored is the evolution of elasticity (e.g. the shear modulus $G$) when boundaries are pinned at a distance $L$. For $L>l^*$, the system remains floppy and $G=0$. For $L<<l_c$, one expects a mean-field argument to apply: pinning boundaries is equivalent to adding springs, leading to an increase of coordination $\Delta z\sim 1/L$. For $z>z_c$ it is known that  $G\sim z-z_c$,\cite{revue} and thus we expect $G\sim 1/L$. The behavior of $G$ at intermediate length $l^*>>L>>l_c$ remains to be explored.

Finally we compare our results with previous works in amorphous solids  made of soft repulsive particles, for which $z>z_c$. Near the unjamming transition where pressure vanishes the  coordination approaches the Maxwell threshold from above ($z\rightarrow z_c$). A vanishing frequency scale $\omega^*\sim z-z_c$ was predicted to characterize the low-frequency part of the spectrum,\cite{Wyart05,Wyart052} as confirmed numerically.\cite{Silbert05}  The same theoretical argument \cite{Wyart05} indicated that boundaries affect elasticity on a length scale $l^*\sim 1/(z-z_c)$. It was later argued,\cite{respprl} based on numerical observations, that $l^*$ characterizes the response to a point force in packings of particles.   Another length scale $l_c$ was observed numerically to characterize the response at a frequency $\omega^*$ \cite{Silbert05} and to affect transport,\cite{vitelli2010} behaviors that were well-captured by effective medium.\cite{wyart2010} Our work extends these results to floppy materials with $z<z_c$, where these lengths and frequency scales characterize the phonon gap and the localization of floppy modes.   However the comparison underlines  an important  disparity: for floppy networks we find both numerically and theoretically  that $l_c$ characterize the response to a local force, whereas $l^*$ appears to characterize the response to a point force in packings.\cite{respprl} More numerical and theoretical investigations are needed to understand this difference.

Acknowledgment: We  thank Y. Elmatad, C. Falc\'on, P. Rowghanian,  A. Grosberg and P. Chaikin for comments on the manuscript.  This work has been supported by the Sloan Fellowship, NSF DMR-1105387, and Petroleum Research Fund \#52031-DNI9. This work was also supported partially by the MRSEC Program of the National Science Foundation under Award Number DMR-0820341.

\footnotesize{
\bibliography{}

}

\end{document}